\documentclass[journal=jpccck,manuscript=article]{achemso}

\usepackage[version=3]{mhchem} 
\usepackage[T1]{fontenc}       

\newcommand{\SNRO}{Sr$_{2}$NiRuO$_4$}

\author{Ke Yang}
\affiliation{Laboratory for Computational Physical Sciences (MOE),
 State Key Laboratory of Surface Physics, and Department of Physics,
  Fudan University, Shanghai 200433, China}

\author{Feng-Ren Fan}
\affiliation{Laboratory for Computational Physical Sciences (MOE),
 State Key Laboratory of Surface Physics, and Department of Physics,
  Fudan University, Shanghai 200433, China}

\author{Alessandro Stroppa}
\affiliation{CNR-SPIN,
Via Vetoio 10, 67100 Coppito, L'Aquila, Italy}

\author{Hua Wu}
\affiliation{Laboratory for Computational Physical Sciences (MOE),
 State Key Laboratory of Surface Physics, and Department of Physics,
  Fudan University, Shanghai 200433, China}
\alsoaffiliation{Collaborative Innovation Center of Advanced Microstructures,
 Nanjing 210093, China}
\email{wuh@fudan.edu.cn}

\title[]
  {Possible High {\it T}$_C$ Layered Ferromagnetic Insulator {\SNRO}: An {\it ab~initio} Study}

\begin{document}

%
%
%
%
%

\begin{abstract}
Magnetic insulators are often antiferromagnetic (AFM) and layered
 AFM compounds usually show low ordering temperature.  On the other hand,  layered  ferromagnetic (FM) insulators with high-$T_{\rm C}$  are very rare although they  could be quite useful for  spintronic applications. Here, using crystal field level analysis in combination with density functional theory calculations as well as Monte Carlo simulations,  we predict
that  the layered insulator {\SNRO} would have a strong FM coupling with   $T_{\rm C}$ as high as  240 K. The tetragonal crystal field in the Ni-O-Ru square plane stabilizes the S=1/2 Ni$^{+}$ and S=3/2 Ru$^{3+}$ states. The unique level ordering and occupation optimize the FM Ni-Ru superexchange  interactions in the checkerboard arrangement, thus suggesting {\SNRO} as
an unusual high-$T_{\rm C}$ layered FM insulator.
 This work highlights the potential of charge-spin-orbital degrees of freedom for stabilizing strong FM coupling in layered oxides.
\end{abstract}

\section{1 Introduction}
Magnetic interactions between magnetic ions bonded by non-magnetic ligands arise via the superexchange mechanism first proposed by Anderson.\cite{Anderson} A set of rules to predict the type of the interactions was proposed, independently, by Goodenough\cite{Goodenough58} and Kanamori,\cite{Kanamori} and they have been  rationalized  as Goodenough-Kanamori-Anderson rules.~\cite{goodenough} Insulating transition-metal oxides are often AFM due to the superexchange interactions.
In  correlated transition-metal oxides, the  strength of the  coupling is given  approximately by $t^2/U$, where $t$  is an effective hopping parameter of the $d$ electrons (hopping between two neighboring transition-metal sites) and
$U$  is the Coulomb interaction between two electrons on a same site.~\cite{goodenough}
Among magnetic insulating oxides, the ferromagnetic (FM) ones are a minority and they are generally connected to   orbital physics (\textit{e.g.}, orbital ordering).~\cite{imada_1998,tokura_2000,dagotto_2005,khomskii} The strength of the FM coupling is approximately given by $t^2J_{\rm H}/U^2$ where $J_{\rm H}$ is
the Hund coupling energy, and this contains a reduction factor $J_{\rm H}/U$ compared with the AFM case.
Since the magnetic ordering temperature is proportional to the exchange couplings, the  FM insulators usually have a much lower ordering temperature than the AFM ones due the presence of $J_{\rm H}/U$ which is about 1/10 - 1/5.~\cite{khomskii}

Bulk FM insulators are interesting for   spintronic applications since they could support    proximity exchange fields in heterostructures and could provide  spin-polarized carriers.\cite{mcguire_2015,huang_2017,gong_2017b,fei2018}
However, a high-$T_{\rm C}$ FM insulator is rare
 due to the aforementioned weakened FM superexchange interactions, and this is even worse in two-dimensional FM materials.\cite{mcguire_2015,hongbo_2016,huang_2017,gong_2017b}
Transition-metal oxides are usually correlated electron systems which offer many  multifunctional properties due to their lattice-charge-spin-orbital degrees of freedom and their cross-couplings.~\cite{imada_1998,tokura_2000,dagotto_2005,khomskii,ou_2015,
xuedong_2016} These materials may host   superconductivity, colossal magnetoresistance, and multiferroicity for example.\cite{imada_1998,tokura_2000,dagotto_2005,khomskii}
In this work, using a combination of crystal field level analyses, density functional calculations, and Monte Carlo simulations,
we study  the interplay among  lattice-charge-spin-orbital degrees of freedoms and we predict a surprisingly strong FM  layered insulator {\SNRO} (see Fig. 1).

\begin{figure}[t]
\centerline{\includegraphics[width=6cm]{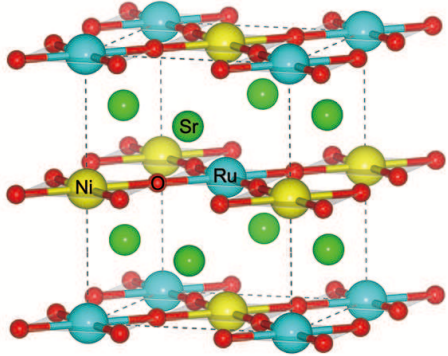}}
\caption{Crystal structure assumed by layered {\SNRO} with the Ni-O-Ru square plane.
}
  \label{fig:st}
\end{figure}

Recently, layered ABO$_{2}$ oxides were prepared via topochemical reduction from ABO$_{3}$ perovskites,\cite{tsujimoto_2007}  and novel properties associated with this particular structure emerge. The representative SrFeO$_{2}$ has already attracted a lot of attention for  its above room-temperature AFM order despite its layered structure.\cite{tsujimoto_2007,xiang_2008} It may undergo a spin crossover and a metal-insulator  transition under pressure, thus forming  an interesting FM half metallic state.~\cite{kawakami_2009} Very recently, the Ni-Ru ordered LaSrNiRuO$_{4}$ was synthesized  and
found to be a FM layered material with $T_{\rm C} \sim$ 200 K.~\cite{patino_2016}
The FM order has been  explained by considering the  superexchange between the unusual low-valence Ni$^+$ and Ru$^{2+}$ cations.\cite{Zhu_2017} The Ni$^+$ ($3d^9$) is in the $S$=1/2 state and has an active $x^2-y^2$ orbital under the influence of a  square planar crystal field. The Ru$^{2+}$ ($4d^6$) shows a $S$=1 state stabilized by a  moderate Hund exchange along with a weak crystal field.\cite{Zhu_2017} If the Ru$^{2+}$ ion were in the nonmagnetic $S$=0 state as suggested in Ref.~\cite{patino_2016}, the magnetic coupling would occur only in the diluted $S$=1/2 Ni$^+$ sublattice, and it would thus be only weakly AFM due to a superexchange between the well separated Ni$^{+}$ ions with the half-filled $x^2-y^2$ orbital. Clearly, this would contradict the measured $T_{\rm C}$ $\sim$ 200 K.\cite{patino_2016}

Starting from these considerations, here we predict a surprisingly strong FM in the layered insulator {\SNRO}. This material would be very similar to the real LaSrNiRuO$_4$,\cite{patino_2016} where half of the Sr atoms are substituted by La atoms. Its structural stability is confirmed by  phonon calculations.
The proposed {\SNRO} compound should have a unique charge-spin-orbital state such as S=1/2 Ni$^{+}$ and S=3/2 Ru$^{3+}$ in the tetragonal crystal field. Such a unique state favors an optimal and maximal superexchange interaction as confirmed by our density functional calculations, and in turn, yields $T_{\rm C}$  $>$ 240 K as predicted by our Monte Carlo simulations.  Therefore, {\SNRO} would be an unusual high-$T_{\rm C}$ layered FM insulator if synthesized.

\section{2 Methods and computational details}
We perform density functional theory (DFT) calculations using the Vienna Ab initio Simulation Package (VASP).~\cite{kresse_1996} The phonon spectrum was   calculated using  phonopy software interfaced with VASP.~\cite{phonopy}
The magnetic properties were studied using DFT, crystal field level diagram, as well as  Monte Carlo simulations.
We start our DFT calculations using the Local-Spin-Density Approximation (LSDA). The ionic potentials including the effect of core electrons are described by the projector augmented wave method.~\cite{kresse_1999}
The plane waves with the kinetic energy up to 500 eV have been employed to expand the electronic wave functions.
A 16-atom unit cell of the Ni-Ru ordered {\SNRO} (space group $I4/mmm$, see Fig.~\ref{fig:st}) has been used to simulate the three magnetic structures studied below.
The integration over the first Brillouin zone is carried out using Monkhorst-Pack grid of $5\times5\times5$ \textit{k}-point mesh.
The structural relaxations are performed till the Hellmann-Feynman force
 on each atom is smaller than 0.01 eV/\AA.
For the phonon calculations, the finite-displacements method is used, with the $2\times2\times2$ supercell (128~atoms) and
 the $3\times3\times2$ \textit{k}-point mesh.
To describe the on-site electron correlation
and to estimate the magnetic exchange parameters,
 the LSDA plus Hubbard U (LSDA+U) method is employed,~\cite{vladimir_1997}
 with the Coulomb and exchange parameters chosen to be
 the common values of U = 6 eV (3 eV) and J = 1 eV (0.6 eV) for Ni 3d (Ru 4d) electrons.~\cite{patino_2016}
Other reasonable values~\cite{ou_2015} of U = 5 eV (2 eV) for Ni 3d (Ru 4d) are also
considered in order to verify that our conclusions remain
 qualitatively unchanged. Moreover, hybrid functional calculations using the Heyd-Scuseria-Ernzerhof (HSE) functional\cite{HSE} are carried out, and the obtained results further confirm our conclusions.

\section{3 Results and discussion}

\subsection{3.1 Phonon calculation}
\begin{figure}[t]
\centerline{\includegraphics[width=7cm]{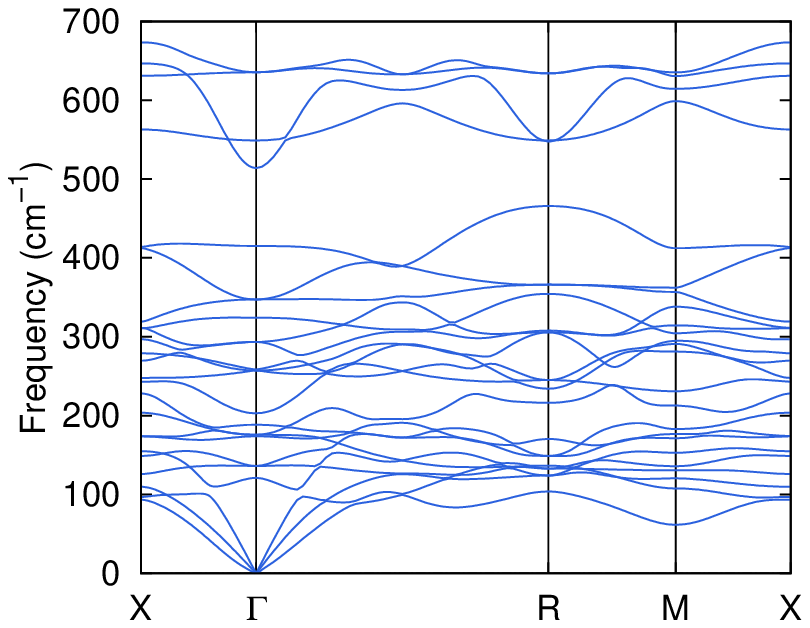}}
 \caption{Calculated phonon spectrum of  {\SNRO}.}
\label{fig:ph}
\end{figure}

We have used the LSDA calculations to optimize the lattice parameters of {\SNRO} in the $I4/mmm$ structure. The obtained lattice constants are $a$=5.48 \AA~ and $c$=6.88 \AA, being well comparable with  the experimental ones of $a$=5.66 \AA~ and $c$=6.90 \AA~ for the parent compound LaSrNiRuO$_{4}$.\cite{patino_2016} To verify the structural stability of {\SNRO}, we have performed the phonon calculations. The obtained phonon spectrum is shown in Fig.~\ref{fig:ph}. There are no imaginary frequencies over the Brillouin zone, and thus {\SNRO}  would be dynamically stable in the $I4/mmm$ structure. As shown below, this stable lattice structure can host a unique charge-spin-orbital state of the Ni and Ru atoms and it is free of a Jahn-Teller distortion. 

\subsection{3.2 LSDA calculations: charge-spin-orbital state and magnetism}

We have studied, for the stable $I4/mmm$ lattice, three different magnetic structures, \textit{i.e.}, FM, A-AF($ab$-intralayer FM, $c$-interlayer AFM) and C-AF ($ab$-intralayer AFM, $c$-interlayer FM). Our calculations show that the FM state is the most stable magnetic configuration  while the  C-AF state is unstable and converges to the metastable A-AF state, see Table~\ref{tb1}. The absence of C-AF state implies that the intralayer AFM coupling is highly unfavorable, and in turn, the intralayer FM is more  stable as found in FM and A-AF states. Both the FM and A-AF states have comparable local spin moments. The spin moment of about 0.8-0.9 $\mu_{\rm B}$/Ni would suggest to a Ni$^{+}$ (3d$^{9}$) S=1/2 state, and that of 1.6-1.7 $\mu_{\rm B}$/Ru would suggest an intermediate-spin Ru$^{3+}$ (4d$^{5}$) S=3/2 state. The large reduction of Ru spin moment from the formal 3 $\mu_{\rm B}$ is due to a strong covalency of the expanded  Ru 4d orbitals with the  oxygen ligands. Note, however, that it is hard to reconcile these calculated spin moments with Ni$^{2+}$ (low-spin S=0 or high-spin S=1) and Ru$^{3+}$ (either low-spin S=1/2 or high-spin S=5/2). In addition, the calculated total spin moment of 3.53 $\mu_{\rm B}$/fu (formula unit) in the FM ground state, including the  contributions from oxygens (0.18 $\mu_{\rm B}$/O) and the interstitial regions, agrees with the Ni$^{+}$ S=1/2 and Ru$^{3+}$ S=3/2 state that has the ideal total spin moment of 4 $\mu_{\rm B}$/fu. This unusual low-valence state will be further verified below. Therefore, despite the similar structure between {\SNRO} and LaSrNiRuO$_{4}$,\cite{patino_2016} they seem to have quite different electronic states. In particular, while for the latter the $S$=1/2 Ni$^{+}$/$S$=0 Ru$^{2+}$ state was suggested,\cite{patino_2016} for the former we propose the $S$=1/2 Ni$^{+}$/$S$=3/2 Ru$^{3+}$ state (but not the $S$=0 Ni$^{2+}$/$S$=0 Ru$^{2+}$ or $S$=0 Ni$^{2+}$/$S$=1 Ru$^{2+}$, see more below). We will show that this unique charg-spin-orbital state  is responsible for the enhancement of the FM couplings  in  {\SNRO} as compared to the parent compound LaSrNiRuO$_{4}$.

\begin{table}[t]
  \caption{Relative total energies $\Delta$E (meV/fu), total and local spin moments
  ($\mu_{\rm B}$) of the FM, A-AF, and C-AF states calculated by LSDA, LSDA+U, and the HSE hybrid functional.
  Sr$_2$NiRuO$_4$ is in the Ni$^{+}$ S=1/2 and Ru$^{3+}$ S=3/2 coupled FM ground state.
  The fixed-spin-moment (FSM) calculations are also included for the total S=1 and S=0 states.
  U$_1$ refers to U = 6 eV (3 eV) for Ni 3d (Ru 4d) electrons, U$_2$ stands for
  U = 5 eV (2 eV) for Ni 3d (Ru 4d) electrons.}
  \label{tb1}
  \begin{tabular}{l@{\hskip3mm}c@{\hskip3mm}c@{\hskip3mm}c@{\hskip3mm}c@{\hskip3mm}c}
\hline
   Sr$_{2}$NiRuO$_4$ & State & $\Delta$E & Total & Ni & Ru\\ \hline
    LSDA   & FM  & 0 & 3.53 & 0.90 & 1.74  \\
           & A-AF   & 18 & 0.00 & 0.82 & 1.65  \\
           & C-AF   & $\longrightarrow$A-AF  \\
           & FSM S=1  & 124 & 2 & 0.17 & 1.28  \\
           & FSM S=0 & 300 & 0 & 0 & 0  \\ \hline
    LSDA+U$_{1}$   & FM    & 0 & 4.00 & 1.13 & 2.09  \\
           & A-AF      & 20 & 0 & 1.14 & 2.06  \\
           & C-AF   &162 & 0 & 0.74 & 1.95  \\ \hline
    LSDA+U$_{2}$   & FM    & 0 & 4.00 & 1.17 & 1.97  \\
           & A-AF      & 27 & 0 & 1.15 & 1.92  \\
           & C-AF   &194 & 0 & 0.68 & 1.81  \\ \hline
    HSE   & FM    & 0 & 4.00 & 1.13 & 2.16  \\
           & A-AF  & 24 & 0 & 1.12 & 2.15  \\
           & C-AF   &299 & 0 & 0.76 & 2.05  \\
\hline
 \end{tabular}
\end{table}

\begin{figure}[t]
\centerline{\includegraphics[width=7.5cm]{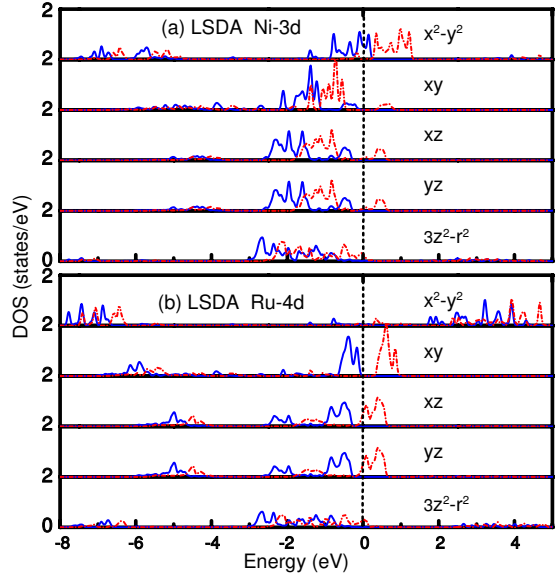}}
 \caption{(a) Ni 3d and (b) Ru 4d DOS of {\SNRO} calculated by LSDA. The blue (red) curves stand for the up (down) spin channel. Fermi level is set at zero energy.
}
  \label{fig:dos_lda}
\end{figure}

We plot in Fig.~\ref{fig:dos_lda} the orbitally resolved density of states (DOS) for the FM ground state. For Ni 3d states, 3z$^{2}$--r$^{2}$ orbital lies at the lowest energy, which is followed by the degenerate xz/yz. The xy orbital can be found at  higher energy  and the
 x$^{2}$--y$^{2}$ orbital highest. Except for the down-spin x$^{2}$--y$^{2}$ state, all other Ni 3d states are (almost) fully occupied, thus suggesting the Ni$^{+}$ (3d$^{9}$) S=1/2 state. As seen in Fig \ref{fig:dos_lda}, Ru 4d states generally lie at higher energy than Ni 3d ones. This is in accordance with the chemical trend that Ru has a higher chemical potential (on site energy) than Fe (both elements in the same group) and Fe higher than the later transition metal Ni. In {\SNRO}, Ru should have a higher valence state than Ni, and the Ni$^{+}$-Ru$^{3+}$ charge state should  be more stable than  Ni$^{2+}$-Ru$^{2+}$ configuration, as seen below. The Ru 4d states have the same energy level sequence as  the Ni 3d ones, but the crystal field splitting of the former is much bigger, as one can see from Fig.~\ref{fig:dos_lda}. In particular, Ru 4d x$^{2}$--y$^{2}$ orbital has a large bonding-antibonding splitting of about 10 eV due to the significant Ru-O hybridization in the $ab$ plane. As a result, the antibonding x$^{2}$--y$^{2}$ state can be found at very high energy
 ($>$ 2 eV above Fermi level) and therefore  it is completely unoccupied. In addition, the down-spin xy and xz/yz states are (almost) empty. This suggests that, besides  the Ni$^+$ S=1/2 state in {\SNRO}, Ru has an electronic configuration  (3z$^{2}$--r$^{2}$)$^{2}$(xz)$^{1}$(yz)$^{1}$(xy)$^{1}$(x$^{2}$--y$^{2}$)$^{0}$, \textit{i.e.}, the intermediate-spin Ru$^{3+}$ (4d$^{5}$, S=3/2) state.
Note that both the Ni$^+$ and Ru$^{3+}$ ions are in an orbital singlet, and thus an insulating behavior of {\SNRO} is expected. While the present LSDA calculations give a small DOS intensity at the Fermi level due to the tiny overlap between the up-spin Ni$^+$ $x^2-y^2$ and the down-spin Ru$^{3+}$ $xz/yz$ bands (see Fig. 3), the insulating behavior is readily recovered  by introducing a  finite electron correlation as shown below.

\begin{figure}[t]
\centerline{\includegraphics[width=6cm]{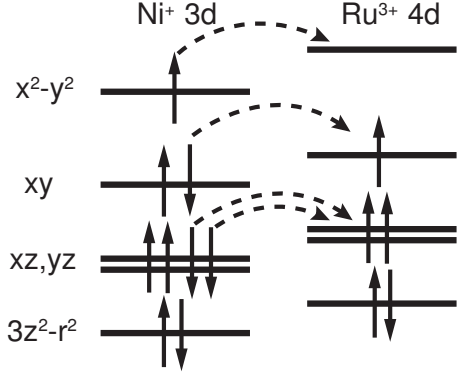}}
 \caption{Schematic crystal field level diagrams of Ni$^{+}$ $S$ = 1/2 and Ru$^{3+}$ $S$ = 3/2. Virtual electron hoppings from Ni$^{+}$ to Ru$^{3+}$ favor FM superexchanges and thus optimize the FM Ni-Ru coupling via the four channels (the dashed curves).
}
  \label{fig:el}
\end{figure}

We plot in Fig.~\ref{fig:el} a schematic  level diagram  for Ni$^+$ S=1/2 and Ru$^{3+}$ S=3/2 ground state as derived  from the above LSDA calculations.  We have also considered different charge-spin states of Sr$_2$NiRuO$_4$, such as Ni$^{2+}$ S=0 and Ru$^{2+}$ S=1 (see Fig.~\ref{fig:el}, without Ni$^{2+}$ $x^2-y^2$ electron and with Ru$^{2+}$ (xz,yz)$^3$), and Ni$^{2+}$ S=0 and Ru$^{2+}$ S=0
(with Ru$^{2+}$ (3z$^{2}$--r$^{2}$)$^{2}$(xz,yz)$^{4}$). These configurations were stabilized  using fixed-spin-moment calculations.
 This approach shows
  that the total S=1 state of Sr$_2$NiRuO$_4$ mimics the Ni$^{2+}$ S=0 and Ru$^{2+}$ S=1 state and turns out to be less stable than the Ni$^{+}$ S=1/2 and Ru$^{3+}$ S=3/2 ground state by 124 meV/fu (see Table~\ref{tb1}). Moreover,
 the total S=0 state (to mimic  the Ni$^{2+}$ S=0 and Ru$^{2+}$ S=0 state) is even higher in energy  than the ground state by 300 meV/fu. Once again, these computational   experiments confirm the Ni$^{+}$ S=1/2 and Ru$^{3+}$ S=3/2 ground state. Since this  electronic state (see Fig.~\ref{fig:el}) is an orbital singlet and it has no orbital degree of freedom, we suggest that {\SNRO} can be stabilized in the $I4/mmm$ structure (as verified by the above phonon calculations) which is free of a symmetry lowering Jahn-Teller lattice distortion. It is important to note that  such unusual charge-spin-orbital states are crucial to produce the  strong FM in this layered insulator, as discussed below.

\subsection{3.3 Ferromagnetic picture}
Now we explain why {\SNRO} should have
a FM magnetic configuration with
a high-$T_{\rm C}$, based on the crystal field pictures and inter-site orbital interactions. By considering   virtual hopping processes, an electronic excitation from Ni$^{+}$ to Ru$^{3+}$ is  possible which gives rise to an  intermediate excited state Ni$^{2+}$/Ru$^{2+}$. Note that  a reverse process would give the highly unbalanced Ni$^{0}$/Ru$^{4+}$ intermediate state and should  be strongly suppressed. Therefore, we expect
 four channels associated with the charge excitation from Ni$^{+}$-Ru$^{3+}$ to Ni$^{2+}$-Ru$^{2+}$, see Fig.~\ref{fig:el}. The
 spin-up x$^{2}$--y$^{2}$ electron would yield a strong in-plane FM coupling via the significant pd$\sigma$ hybridizations of the  Ni-O-Ru bonds. The spin-down xy electron would enhance the in-plane FM coupling via the relatively weaker pd$\pi$ hybridizations (but with a smaller charge excitation gap of less than 1 eV, see Fig.~\ref{fig:dos_lda}), and this is also the case for the down-spin xz/yz electrons. Moreover, the down-spin xz/yz electrons would yield a FM interlayer coupling via the dd$\pi$ hybridizations (albeit there no bridging apical oxygens). Note that the particular Ni$^{+}$ S=1/2 and Ru$^{3+}$ S=3/2 state in the planar tetragonal crystal field is quite unique in producing the FM couplings via the inter-site orbital interactions. Except for the inactive 3z$^{2}$--r$^{2}$ orbital (doubly occupied due to the planar tetragonal crystal field), all the other four orbitals contribute to the FM coupling and therefore give rise to an optimal FM order.

\subsection{3.4 LSDA+U and hybrid functional calculations}
 \begin{figure}[t]
\centerline{\includegraphics[width=7cm]{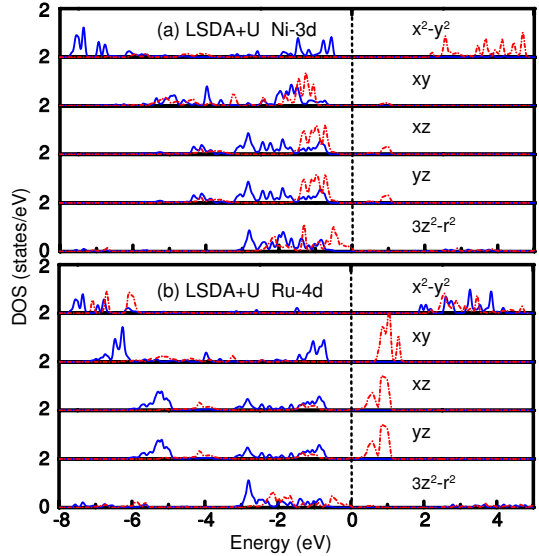}}
 \caption{(a) Ni 3d and (b) Ru 4d DOS of {\SNRO} calculated by LSDA+U$_1$ (see also Table 1).
 The blue (red) curves stand for the up (down) spin channel. Fermi level is set at zero energy.
}
  \label{fig:dos_ldaU}
\end{figure}
To support our picture,  we have performed additional LSDA+U calculations to estimate the strength of both the intralayer and interlayer FM couplings in Sr$_2$NiRuO$_4$ in the stable Ni$^{+}$ S=1/2 and Ru$^{3+}$ S=3/2 state. As seen in Table \ref{tb1}, LSDA+U calculations account for the electron localization and stabilize the different magnetic structures for this layered  insulator. Moreover, using constrained LSDA+U calculations,\cite{ou_2015,xuedong_2016} we have tested the Ni$^{2+}$ S=0 and Ru$^{2+}$ S=1 state, and the Ni$^{2+}$ S=0 and Ru$^{2+}$ S=0 state. The former turns out to be unstable and converges exactly to the Ni$^{+}$ S=1/2 and Ru$^{3+}$ S=3/2 ground state. The latter is much higher in energy than the ground state by 1550 meV/fu. This confirms the above fixed-spin-moment calculations and both conclude the $S$=1/2 Ni$^{+}$/$S$=3/2 Ru$^{3+}$ ground state.
The insulating electronic structure can be seen in Fig.~\ref{fig:dos_ldaU}, showing the orbitally resolved DOS results for the FM ground state given by LSDA+U. Again, the Ni$^{+}$ (S=1/2) and Ru$^{3+}$ (S=3/2) ground state is verified, by counting the spin-orbital occupations of both Ni 3d and Ru 4d states. Therefore, our interpretation in terms of Ni$^{+}$ (S=1/2) and Ru$^{3+}$ (S=3/2) is robust with respect to different computational schemes.

By mapping  the total energy differences between the FM ground state and the C-AF state, and A-AF states to a Heisenberg spin Hamiltonian (see Table~\ref{tb1}), the FM intralayer coupling parameter $J_{ab}$ is estimated to be --27.0 meV (8$J_{ab}$S$_{Ni}$S$_{Ru}$ = --162 meV) and the FM interlayer one $J_{c}$ = --6.7 meV (4$J_{c}$S$_{Ni}$S$_{Ru}$ = --20 meV) when referring to the LSDA+U$_1$ scheme where U = 6 eV (3 eV) could be an upper limit for Ni 3d (Ru 4d) electrons. When we refer to the LSDA+U$_2$ scheme with smaller
U = 5 eV (2 eV) for Ni 3d (Ru 4d) electrons, the FM exchange parameters get even bigger:
$J_{ab}$ = --32.3 meV and $J_{c}$ = --9.0 meV. This is reasonable because by decreasing U parameter, the electron localization decreases and the energy separation between Ni 3d and Ru 4d states becomes smaller, thus favoring the virtual hoppings that we discussed before,
and thus enhancing the FM couplings. Therefore, the estimated $T_{\rm C}$ would  be a minimum based on the LSDA+U$_1$ results. All these results show that indeed the intralayer FM coupling is significant and the interlayer one is also strong,  thus confirming  the above spin-orbital modeling analyses (Fig.~\ref{fig:el}). As far as the local spin moments are concerned,  with LSDA+U$_1$ (and LSDA+U$_2$)
values of about
1.1 $\mu_{\rm B}$/Ni$^{+}$ and 2.0 $\mu_{\rm B}$/Ru$^{3+}$ are larger than the corresponding LSDA ones, due to the electron localization introduced  by the Hubbard U. The LSDA+U values and the calculated total spin moment of 4 $\mu_{\rm B}$/fu for the FM ground state clearly represent the unique Ni$^{+}$ S=1/2 and Ru$^{3+}$ S=3/2 state.

Furthermore, hybrid functional HSE calculations are carried out, and they reach the same conclusion as the LSDA and LSDA+U calculations do, that is, {\SNRO} is stabilized in the $S$=1/2 Ni$^{+}$/$S$=3/2 Ru$^{3+}$ ground state, see Table I. While the $J_c$ value changes insignificantly by a comparison between the HSE and LSDA+U$_1$ (and LSDA+U$_2$) calculations of the FM and A-AF states, the $J_{ab}$ value is nearly doubled when comparing the HSE and LSDA+U$_1$ calculations of the FM and C-AF states. Therefore, our HSE calculations would give a much higher $T_{\rm C}$ than the LSDA+U$_1$ calculations do.

\begin{figure}[t]
\centerline{\includegraphics[width=8cm]{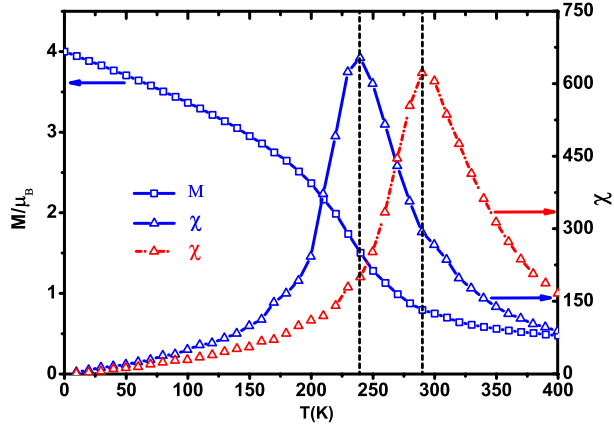}}
 \caption{Monte Carlo simulations of magnetization and magnetic susceptibility of {\SNRO} as a function of temperature. $T_{\rm C}$ is 240 K (blue curves) based on $J_{ab}$ = --27.0 meV and $J_{c}$ = --6.7 meV, and $T_{\rm C}$ is 290 K (red curve) based on $J_{ab}$ = --32.3 meV and $J_{c}$ = --9.0 meV.
}
\label{fig:mc}
\end{figure}

\subsection{3.5 Monte Carlo simulations: high $T_{\rm C}$ layered ferromagnetism}

Finally, we perform Monte Carlo simulations to confirm our prediction of a high  $T_{\rm C}$ for the layered insulator {\SNRO}, using the stable Ni$^{+}$ S=1/2 and Ru$^{3+}$ S=3/2 states and the above LSDA+U$_1$ (and LSDA+U$_2$) calculated FM Ni-Ru exchange parameters. We have used a three dimensional spin lattice and we assumed a Heisenberg spin Hamiltonian for the nearest neighboring Ni-Ru sites. Each spin is isotropic and can be rotated arbitrarily in three dimensional space, and the noncollinearity between the spins has been taken into account. The Metropolis method~\cite{nicholas_1949} was used in the simulations. The system is fully relaxed by 3$\times$10$^{7}$ spin flips
at each temperature. After the system reaches equilibrium at a given temperature, we calculate the magnetization M (the total magnetic moment of the system) and the magnetic susceptibility by $\chi$=($<$M$^{2}$$>$--$<$M$>$$^{2}$)/k$_{\rm B}$ $T$. M is sampled after each 6$\times$10$^{3}$ spin flips, and 6$\times$10$^{3}$ magnetizations are used to take the average. Based on the LSDA+U$_1$ FM exchange parameters $J_{ab}$ = --27.0 meV and $J_{c}$ = --6.7 meV, the simulation results are shown in Fig.~\ref{fig:mc} (blue curves). The calculated magnetization and magnetic susceptibility show that the Curie temperature $T_{\rm C}$ is 240 K, which may be a minimum but is already quite high for a layered insulator. Moreover, when we use $J_{ab}$ = --32.3 meV and $J_{c}$ = --9.0 meV from LSDA+U$_2$, $T_{\rm C}$ is increased up to 290 K, see the red curve in Fig.~\ref{fig:mc}. In addition, as our HSE calculations give a nearly doubled $J_{ab}$ value than the LSDA+U$_1$, the corresponding $T_{\rm C}$ should be significantly enhanced and then should be much higher than room temperature. Therefore, we have predicted an unusual strong FM in the layered insulator {\SNRO}, whose $T_{\rm C}$ is close to or ever higher than room temperature. This prediction of a high $T_{\rm C}$ in the stable {\SNRO} calls for an experimental proof.

\section{4 Conclusion}
In summary, we have predicted a surprising strong FM coupling in the stable layered insulator {\SNRO}, using crystal field level analyses, density functional calculations for phonon and magnetism, and Monte Carlo simulations. Our results show that {\SNRO} has the unique S=1/2 Ni$^{+}$ and S=3/2 Ru$^{3+}$ state in the planar tetragonal crystal field. The unique charge-spin-orbital states give rise to optimal FM couplings, which are confirmed by density functional calculations. Our Monte Carlo simulations show that {\SNRO} should have a  $T_{\rm C}$ $>$ 240 K thus representing a possible  FM layered insulator with surprising high $T_{\rm C}$.  This work highlights how    unusual charge-spin-orbital states could stabilize   strong FM interactions in novel layered oxide insulators. This certainly calls for further theoretical and experimental studies in this direction in order to design
new layered FM materials suitable for device applications.

\begin{acknowledgement}
H.W. conceived and supervised the research.
K.Y. and F.-R.F. contributed equally to the work by carrying out the calculations with assistance from H.W. and A.S.
All authors performed data analysis and co-wrote the manuscript.
 This work was supported by the NSF of China (Grants No. 11474059 and No. 11674064) and the National Key Research and Development Program of China (Grant No. 2016YFA0300700). F.-R. Fan was also partially supported by the Key Laboratory of Polar Materials and Devices (MOE), East China Normal University (Grant No. KFKT20160003).
\end{acknowledgement}

%

\bibliography{Sr2NiRuO4}

 \begin{figure}[h]
\centerline{\includegraphics[width=8.5cm, height=3.9cm]{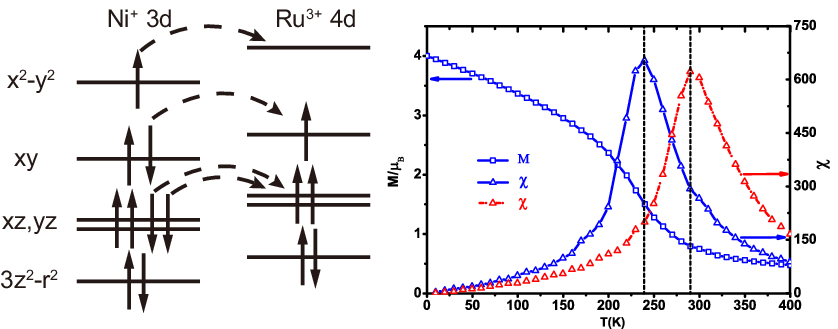}}
  \caption{TOC Graphic
}
\end{figure}

\end{document}